# Detailed description of spontaneous emission

**M. V. Guryev**

Author is not affiliated with any institution. Home address: Shipilovskaya st., 62-1-172, 115682-RU, Moscow, Russian Federation

E-mail: mvguryev@mail.ru

The wave side of wave-photon duality, describing light as an electromagnetic field (EMF), is used in this article. EMF of spontaneous light emission (SE) of laser excited atom is calculated from first principles for the first time. This calculation is done using simple method of atomic quantum electrodynamics. EMF of SE is calculated also for three types of polyatomic light sources excited by laser. It is shown that light radiated by such sources can be coherent, which explains recent experiments on SE of laser excited atoms. Small sources of SE can be superradiant, which also conforms to experiment. Thus SE is shown not to be a random event itself. Random properties of natural light are simply explained as a result of thermal excitation randomness without additional hypotheses. EMF of SE is described by simple complex functions, but not real ones.

## 1. Introduction

Spontaneous emission (SE) of an excited atom is the simplest example of a process, where atom interaction with electromagnetic field (EMF) takes place. The first theory of SE was formulated by Dirac in 1927 and further by Weiskopff in 1930. The modern version of this theory see, e. g., [1, 2]. This theory describes photon emission and is now accepted as a basic one. The main result of this theory is the well known formula for the probability of photon emission event. This formula is validated by numerous experiments but it gives nothing more in addition to the probability of SE. Therefore SE can be considered within photon description only as a random event of photon emission.

Therefore, in turn, one may think that SE of excited atom cannot be phase coherent with respect to an excitation laser. This point of view is often repeated in textbooks on optics and lasers.

But recently it was shown that light radiated by SE of laser excited atoms can be coherent with respect to an excitation laser [3, 4]. These results manifestly contradict any random description of SE. We must solve this contradiction.

It is obvious, that for this purpose we need to obtain more detailed information on SE, than just the value of photon emission probability.

We know that there exists wave-photon duality, that is, photons and waves of EMF can be described as different physical objects, although they are assumed to exist, in the general case, together. We speak usually about "photon interference" and other photonic processes. But if we want to describe interference, diffraction or any other light transformations quantitatively, we have to write wave equation for EMF and work with it.

We must acknowledge that description of light as electromagnetic waves is irreplaceable for quantitative calculations of light transformations without absorption and other problems of classical optics.

On the other part, we can really count photons by observing jumps of photocurrent. Unfortunately, in many cases we have to speak about photons, because we know very little about EMF besides the fact of light existence in our problem. SE is exactly the same case.

I remind the reader that experimenters can measure light intensity, which is proportional to the squared modulus of EMF, in any point of their laboratory. They can also measure frequency, phase shift and vector properties of observable EMF using interference methods. Therefore we must acknowledge that experimenters can measure EMF as a physical object completely. Therefore we should not confine ourselves to the photon description of light emission

EMF is described in the general case by a four variables function. It is obvious that such a function (if we would find it) can describe SE essentially in a more detailed manner than the only number – probability of SE – which we now know from the usual random photon description. Thus we may hope that EMF description of SE can be useful for our goal to explain unexpected experimental results [3, 4].

Of course, waves of EMF are used in quantum electrodynamics (QED) and in other calculations for light description. But in all cases within atomic physics these waves were chosen as the simplest proper solutions of wave equation. These solutions were convenient for calculations but their relation to real EMF remains, strictly speaking, unknown. We can say only that these waves are like real ones if calculations are more or less validated by experiment. So the immediate goal of this article is to calculate EMF of SE strictly to describe the simplest electromagnetic phenomenon.

It is possible to say that the first step in this direction was done in [5]. In this article SE was considered as a process, but not a jump; energy density during SE was obtained "by lengthy calculation". But we know that exactly 4-vector of EMF is a basic quantity for strict description of all electromagnetic phenomena. Therefore calculation of EMF of SE would be the next step in our understanding.

The first order perturbation theory of atomic QED (first order diagram) see, e. g., [6] will be used as a foundation of the calculation method in this article. This simple calculation method gives opportunity of direct analysis of its physical meaning, practical applications and drawbacks. I remind the reader that atomic QED is the most precise theory, developed by humankind.

EMF of SE is calculated for the first time in this article. The rest of the article is organized as follows. In Sec. **2** EMF of SE for a single excited atom is calculated. In Sec. **3** EMF of SE for different polyatomic objects is calculated and analyzed. Sec. **4** contains discussion and conclusions. Discussions of special cases of polyatomic sources are situated in corresponding subsections.

## 2. The EMF of a single radiating atom

Any EMF is a solution of the d'Alembert equation (DE) for 4-vector-potential see, e. g., [7]. We won't use more complicated Maxwell equations, because, strictly, they can be considered as consequences of simpler DE, which is usually named as the "wave equation for 4-vector-potential". Green's functions of DE (photon propagators) are an essential part of QED. By way of a more general formulation of our problem we'll use 4-dimensional methods in the very beginning of this section.

We'll try for EMF of SE as a solution of DE with 4-current of the source in the right side, that is, with source of EMF.

We'll use atomic units. Therefore DE in our case has the form

$$\frac{\partial^2 A}{c^2 \partial t^2} - \frac{\partial^2 A}{\partial x^2} - \frac{\partial^2 A}{\partial y^2} - \frac{\partial^2 A}{\partial z^2} = j \qquad (1)$$

where $A$ is 4-vector potential describing EMF and $j$ is 4-current created by EMF source. We use the first order of the perturbation theory and assume the condition for Eq. (1) solution: $A = 0$ in the entire 3-space if $t = 0$. Thus we consider free emission of EMF without any outer limitations and complications. First order solution of Eq. (1) in this case is

$$A = \int j D_R \, d^4 x_1 = \frac{1}{(2\pi)^4} \int \exp(ipx) \frac{d^4 p}{p^2 + i0} \int j \exp(-ipx_1) d^4 x_1 \qquad (2)$$

Here $D_R$ is the photon propagator which is used in QED. We use specially the retarding Green function because we describe an irreversible radiation process but not a stationary state of interacting objects where propagator $D_c$ is in use within atomic QED. In second Eq. (2) $D_R$ is written in integral representation to compare our intermediate result with well known ones.

The last integral in second Eq. (2) is the usual within QED vectorial matrix element for probability of SE [6]. Second Eq. (2) implies that our 4-vector potential $A$ is a result of the superposition of all possible plane waves which represent a complete system of functions and the mentioned matrix element is the weight factor. Thus Eq. (2) is a direct generalization of the usual method used in QED for the calculation of SE probability.

Let's note that Eqs. (1) and (2) are valid for SE of any excited objects for which we can write the expression for 4-current and for which the perturbation theory is applicable. Our method can be applied, e. g., to cyclotron radiation. In this article we'll be limited to atom SE.

We will use for simplicity spinless nonrelativistic atom wave functions with one electron, hence 4-current $j$ can be written within QED approximately as

$$j = -i\psi_2(\mathbf{r}_1) p \psi_1(\mathbf{r}_1) \exp(i(\omega_2 - \omega_1)t_1) \qquad (3)$$

where $p = (0, -i\nabla)$ is 4-momentum operator within the usual nonrelativistic approximation [6]; $\psi_1(\mathbf{r}_1)\exp(-i\omega_1 t_1)$ and $\psi_2(\mathbf{r}_1)\exp(-i\omega_2 t_1)$ are initial and final atom states respectively.

In our case the time component of 4-current $j$ is equal to null. Therefore we change $A \to \mathbf{A}$ and work further only with 3-dimensional vectors.

We consider only two levels in our problem and accept for brevity $\omega_2 = 0, \omega_1 \to \omega$. We introduce also the quantity $k \equiv \omega/c$. It is necessary to take into account line width (finite mean lifetime) of the initial state, hence we assume that this state frequency is $\omega = \omega_0 - i\gamma$, where is $2\gamma\tau_E = 1$ and $\tau_E$ mean lifetime of the initial state of atom. This implies that parameter $k$ is complex and the excited state of the atom was prepared in moment $t_1 = 0$.

We assume in this section that atom nucleus is situated at the origin of the coordinates.

All the simplifications we made are not essential for the main results of the article because the Eq. (3) and all following calculations can be easily generalized to any quantum object with necessary precision.

We use Green function in Feynman gauge [1, 5, 6]

$$D_R = \frac{1}{c^2 |\mathbf{r} - \mathbf{r}_1|} \delta\left(\frac{|\mathbf{r} - \mathbf{r}_1|}{c} - (t - t_1)\right) \delta_{\mu\nu} \quad (4)$$

Here $\delta_{\mu\nu}$ is the metrical tensor. We substitute the Eqs. (3) and (4) into Eq. (2), calculate elemental time integral with $\delta$ – function within $0 < t_1 < t$ and get

$$\mathbf{A} = -i \int \frac{\psi_2(\mathbf{r}_1) \nabla \psi_1(\mathbf{r}_1) \exp(i(k|\mathbf{r} - \mathbf{r}_1| - \omega t)) d\mathbf{r}_1}{c |\mathbf{r} - \mathbf{r}_1|}, \, ct > |\mathbf{r} - \mathbf{r}_1| \quad (5)$$

We limit ourselves to consideration of EMF on large distances, that is, we neglect atom dimensions, accept condition $r \gg r_1$ and replace $r \to R$. Condition $ct > r$ is strict requirement of causality revealed before in [5].

We introduce vector $\mathbf{k} \equiv \mathbf{p}k$ where $\mathbf{p} \equiv \mathbf{R}/R$. Vector $\mathbf{R}$ should be interpreted as the radius-vector of the point where we observe EMF. Of course we assume that EMF can be observed and measured with modern technique without substantial EMF disturbance in any point of space. Vector $\mathbf{r}_1$ should be considered as the radius-vector of atomic electron.

We limit our consideration to the dipole approximation. It is worth noting that this approximation is not obligatory and higher approximations can be easily calculated.

We neglect the quantity $\mathbf{kr}_1$; replace gradient operator on $\mathbf{r}_1$ operator using the well known equality $<2|\nabla|1> = (E_1 - E_2)<2|\mathbf{r}_1|1>$ and get

$$\mathbf{A} = -\frac{ik}{R} \exp(i(kR - \omega t)) <2|\mathbf{r}_1|1>, \, \mathbf{A} = 0 \text{ if } ct < R \quad (6)$$

Eq. (6) represents the solution to our problem for a single atom.

Below we transform Eq. (6) into traditional form of solutions of Maxwell equations for comparison with the usual description of light waves. For this transformation we use the known formulas of classical electrodynamics for the strength of electric and magnetic fields, expressed in terms of 4-dimentional quantities see, e.g., [8]. We get directly

$$\mathbf{E} = -\frac{k^2}{R} <2|\mathbf{r}_1|1> \exp(i(kR-\omega t)), \ \mathbf{E} = 0 \text{ if } ct < R \qquad (7)$$

$$\mathbf{H} = \frac{k^2}{R} <2|\mathbf{n}\times\mathbf{r}_1|1> \exp(i(kR-\omega t)), \ \mathbf{H} = 0 \text{ if } ct < R \qquad (8)$$

It is useful to consider the modulus of $\mathbf{A}$ which is determined by Eq. (9) as

$$A = \frac{\omega_0}{cR} \exp\left(-\gamma\left(t-\frac{R}{c}\right)\right), \ A = 0 \text{ if } ct < R \qquad (9)$$

We can see from Eq. (9) that the value of $A$ is maximal if $R = ct$, that is, at the boundary of the EMF. This value decreases with distance as $1/R$. Such decreasing is typical for EMF far enough from the source. The value of $A$ decreases exponentially with time in the atom where $R = 0$. It is possible to say that EMF gradually leaves the atom with mean EMF lifetime $\tau_F \equiv 1/\gamma$. This lifetime will be used in the subsection devoted to natural light. Let's recollect that experimental mean lifetime of an excited atom is $\tau_E = 1/2\gamma$.

We can see from Eqs. (6) and (9) that emitted EMF propagates as a diverging spherical wave packet. If $\gamma t \gg 1$, this wave packet can be considered to be a diverging spherical layer with "thickness" $l \approx (c/\gamma) = 2c\tau$. Our wave packet moves with light velocity and can be interpreted as a photon, as well as wave packets in quantum mechanics, which can be interpreted as classically moving particles [9].

### 3. The EMF of polyatomic light sources

We'll consider here light sources containing $N$ non-interacting identical excited atoms situated in different positions inside the source. Currents of non-interacting atoms are assumed to be additive. We neglect light absorption, scattering, stimulated radiation and reemission in our consideration of SE. We assume that all the atoms were excited by laser instantly but non-simultaneously, because laser pulse moves with light velocity. Then we can write the sum EMF of SE of a polyatomic source in the general case as

$$\mathbf{A}_S = \sum_n \mathbf{A}_n \equiv -\frac{ik}{R}\sum_n \exp\left(i\left(k|\mathbf{R}-\mathbf{r}_n|-\omega(t-t_n)\right)\right) <2|\mathbf{r}_1-\mathbf{r}_n|1> \quad (10)$$
$$\mathbf{A}_n = 0, \text{ if } c(t-t_n) < |\mathbf{R}-\mathbf{r}_n|$$

where $n$ is the atom number, $t_n$ is the moment of this atom excitation, $\mathbf{r}_n$ the radius-vector of its nucleus and $\mathbf{r}_1$ is the integration variable in every matrix element. If our source is excited by a single short laser pulse, it is possible to write $ct_n = \mathbf{r}_n\mathbf{m}$ where $\mathbf{m}$ is the unit vector of the laser pulse direction. In this case we can write a simplification for Eq. (10): $\omega t_n = k\mathbf{m}\mathbf{r}_n$.

We'll assume that $R \gg r_n$, thus we are limited to source dimensions small enough in comparison with the distance to the EMF detector. Therefore the general factor $1/|\mathbf{R}-\mathbf{r}_n|$ is *a priori* written in Eq. (10) approximately as $1/R$. This approximation will be in use below on default everywhere besides the exponent index.

Atom wave functions describing states $|1>$ and $<2|$ should be written in Eq. (10) for quantitative calculations as functions of vectors $(\mathbf{r}_1 - \mathbf{r}_n)$.

Orientations of transition dipole moments for all identical atoms are the same if atoms were excited by light with definite linear polarization. In this case all atom matrix elements are equal and we can factor out one.

This implies that a linearly polarized laser pump creates always light of SE with definite polarization, if all mentioned above conditions are really satisfied.

### 3. 1. *Small source superradiance*

Firstly we consider the simplest extreme case for a small source and name it source1. We'll suppose in this case that $kr_n \ll 1$ for all $r_n$. Then all source1 dimensions are much less than wave length and we neglect $r_n$ in all exponent indexes.

On these conditions all summands in Eq. (10) turn out to be approximately equal and we can write in this case

$$A_S = -\frac{ikN}{R}\exp\left(i\left(kR - \omega t\right)\right) < 2 | \mathbf{r}_1 | 1 > \qquad (11)$$

Eq. (11) implies that light of source1 is coherent and light intensity $I \propto N^2$. This is a well known sign of superradiance; see e. g., [10] and references therein. For brevity I have named such kind of emission also superradiance.

I draw the reader's attention to the lack of any hypotheses about cooperative states in our derivation, although the usual theory of superradiance requires them [11].

We can say that constructive interference, and superradiance as its consequence, takes place always for any instantly and coherently excited light source, which is small enough independently of its form and other details. This result conforms to the majority of experimental results, see e. g. [10] and references therein.

### 3. 2. *More complex sources*

Now we move to more complex cases when the type of light and its intensity depends on details of the source's form. Here we'll suppose that the more soft condition $kr_n^2 \ll R$ is satisfied. Let's note that this condition can be satisfied for any source, if it is situated far enough from the detector. We expand the coordinate term in the exponent index in series and get approximately

$$\mathbf{A}_S = -\frac{ik}{R}\exp(-i(kr+\omega t))<2|\mathbf{r}_1|1>\sum_n \exp i(\mathbf{kr}_n+\omega t_n) \qquad (12)$$

The Eq. (12) describes radiation of SE, which is coherent independently of excited atoms positions and moments of excitation, because the last factor of Eq. (12) introduces only a constant and has no effect on alternating phase of light wave. Thus it determines only light intensity. This factor, of course, depends on the dimensions and form of the source.

If excited atoms are located in the source randomly, we'll name it source2. For such sources we get directly $I \propto N$. A more interesting case is possible.

Let's suppose that atoms of this source, which we'll name source3, are situated periodically, so that $k(\mathbf{p}+\mathbf{m})(\mathbf{r}_n - \mathbf{r}_{n\pm 1}) \approx 2\pi a$ for all values $n$. Here $a$ is an arbitrary integer. For such source we have $\sum_n \exp(i\mathbf{kr}_n) \propto N$ and as a result $I \propto N^2$. This implies that in this case we can also obtain superradiance.

### 3. 3. *The EMF of natural (thermal) light*

The temperature dependence of thermal light intensity (Planck distribution) is well described by methods of statistical physics. Hence we'll have to explain natural light properties as consequences of conditions of atomic excitation, corresponding to the statistical description. Within statistical physics we should assume every atom in the source to be excited independently and randomly, that is, in different random moments. In this case we can say that our source consists of $N$ independent sources which have nothing connecting them.

Hence we have to assume that in a natural source $t_n$ and $r_n$ in Eq. (10) are random and $t_n$ is a "smeared" quantity. Orientation of the quantization axis is also random and transition dipole moments are random too. In this case we cannot factor out one of them, and the limitation $R \gg r_n$ is not useful. Therefore a natural light cannot have any definite polarization vector and is observed as unpolarized.

Eq. (10) with the mentioned random conditions describes EMF of a natural (thermal) light of a polyatomic source with any dimensions. Observable light of such a source is, of course, incoherent with laser light.

It has been well known for a long time that we can split natural light into two beams and study their mutual coherence. This property of thermal light quite naturally follows from Eq. (10). It is obvious that $c\tau_F$ and $\tau_F$ are coherence length and coherence time respectively, if the beam of natural light is combined with a copy of itself. We have to use here exactly $\tau_F$ but not $\tau_E$ which is mean lifetime of excitation in an atom. This prediction can be checked by experimenters.

We have obtained an important result. The main properties of natural light are described as ones created by randomness of excitation conditions, but not by randomness of atom SE itself. It is possible to say that photon description imposed on us random description of SE.

Collisions in plasma are also usually random, if the plasma is not specially "organized". Therefore gas discharge light is often random and is like a thermal one.

### 4. Discussion and conclusions

The first result of the article is obvious: vector properties of EMF are determined by source properties.

The second result of the article is obvious too because the Eqs. (6), (7) and (8) are complex, but not real. This fact implies that, in general, vector-potential describing electromagnetic waves, is a complex function, as well as wave functions in quantum mechanics. This result conforms well to QED because Green functions of DE are constructed as superpositions of products of complex plane waves, but not real ones [7]. The reader can ascertain this fact by looking at the second Eq. (2) in this article.

At first glance existence of complex EMF contradicts the generally accepted opinion: vector-potential is real. The fact of the matter is that expressions for 4-vector potencial are usually postulated to be real. It was natural to extrapolate successful methods of classical electrodynamics to the microworld which we describe as a wave world. Such extrapolation created some difficulties which were eliminated by the usage of two real polarizations instead of the usual in experiment one in-plane polarization vector see, e. g., [5, 6].

Let's note that $\mathbf{EH} = 0$. The same equality is often accepted in optics, although the obtained radiated wave described by Eqs. (7) and (8) is not a plane one, but is a diverging spherical wave packet.

Field of SE of a single atom after laser excitation is coherent to laser and obviously does not contain any uncertainty or randomness. This EMF is uniquely determined by Eq. (6). Obtained light is linearly polarized. The direction of the polarization vector coincides with the direction of the dipole moment of atom transition. Thus EMF of SE of a single atom has simple and natural properties.

Light of polyatomic sources, which were excited by linearly polarized laser light, is also linearly polarized, if the detector is far enough from the source. Light of the simplest polyatomic sources is coherent in this case, that explains theoretically the results of [3, 4].

All predicted light properties of polyatomic sources can be observed, only if the mentioned above conditions are fulfilled at least partially and there are no considerable disturbing interactions inside the source.